\documentclass[10pt,onecolumn,twoside,draftclsnofoot]{IEEEtran}%

\usepackage[tbtags]{amsmath} 
\usepackage{amssymb} 
\usepackage{enumerate}

\usepackage{amsfonts}
\usepackage[caption=false]{subfig}
\usepackage{color}
\usepackage{cite}

\usepackage{hyperref}

\usepackage{epsfig, graphicx, psfrag}

\usepackage{wasysym}

\usepackage{cases}

\usepackage{scrextend}

\usepackage{tikz}
\usetikzlibrary{shapes,arrows,decorations.markings,positioning}

\usetikzlibrary{shapes,arrows,decorations.markings}

\tikzstyle{plant} = [draw, fill=red!5, rectangle, 
    minimum height=3em, minimum width=6em]
\tikzstyle{block} = [draw, fill=blue!5, rectangle, 
    minimum height=3em, minimum width=6em]
\tikzstyle{sum} = [draw, fill=yellow!10, circle, node distance=1cm]
\tikzstyle{coord} = [coordinate]
\tikzstyle{gain} = [draw, fill=red!5, regular polygon, regular polygon sides=3, shape border rotate=-90]
\tikzstyle{pinstyle} = [pin edge={to-,thick,black}]

\tikzstyle{BitPipe} = [thick, decoration={markings,mark=at position
   1 with {\arrow[semithick]{open triangle 60}}},
   double distance=1.4pt, shorten >= 5.5pt,
   preaction = {decorate},
   postaction = {draw,line width=1.4pt, white,shorten >= 4.5pt}]

\usepackage{ifthen}
\providecommand{\VersionLength}{long}
\newcommand{\ver}{\ifthenelse{\equal{\VersionLength}{long}}}
\newcommand{\nver}{\ifthenelse{\equal{\VersionLength}{short}}}

\usepackage[normalem]{ulem}
\usepackage{cancel}

\let\oldsqrt\sqrt
\def\sqrt{\mathpalette\DHLhksqrt}
\def\DHLhksqrt#1#2{%
\setbox0=\hbox{$#1\oldsqrt{#2\,}$}\dimen0=\ht0
\advance\dimen0-0.2\ht0
\setbox2=\hbox{\vrule height\ht0 depth -\dimen0}%
{\box0\lower0.4pt\box2}}

\usepackage{amsthm}
\theoremstyle{plain}
\newtheorem{thm}{Theorem}
\newtheorem{lemma}{Lemma}
\newtheorem{assert}{Assertion}
\newtheorem{corol}{Corollary}
\newtheorem{prop}{Proposition}
\theoremstyle{definition}
\newtheorem{defn}{Definition}
\newtheorem{algo}{Algorithm}

\theoremstyle{remark}
\newtheorem{remark}{Remark}

\providecommand{\thmref}[1]{Thm.~\ref{#1}}

\providecommand{\defnref}[1]{Def.~\ref{#1}}
\providecommand{\secref}[1]{Sec.~\ref{#1}}
\providecommand{\lemref}[1]{Lem.~\ref{#1}}
\providecommand{\propref}[1]{Prop.~\ref{#1}}
\providecommand{\assertref}[1]{Assert.~\ref{#1}}
\providecommand{\remref}[1]{Rem.~\ref{#1}}
\providecommand{\figref}[1]{Fig.~\ref{#1}}

\providecommand{\algoref}[1]{Alg.~\ref{#1}}
\providecommand{\condref}[1]{Cond.~(\ref{#1})}

\providecommand{\tableref}[1]{Table~\ref{#1}}

\newcommand{\ie}{i.e.}
\newcommand{\eg}{e.g.}

\newcommand{\viz}{viz.}
\newcommand{\etal}{et al.}

\newcommand{\reals}{\mathbb{R}}
\newcommand{\ints}{\mathbb{Z}}

\newcommand{\nats}{\mathbb{N}}

\newcommand{\e}{\text{e}}

\newcommand{\mD}{\mathcal{D}}
\newcommand{\mE}{\mathcal{E}}

\newcommand{\cI}{\mathcal{I}}

\newcommand{\Comment}[1]{}
\newcommand{\old}[1]{}
\newcommand{\rem}[1]{}

\newcommand{\supp}[1]{ \mathrm{Supp} \left\{ #1 \right\} }

\newcommand{\hx}{\hat{x}}
\newcommand{\hX}{\hat{X}}

\newcommand{\hW}{\hat{W}}
\newcommand{\hw}{\hat{w}}

\newcommand{\tG}{\tilde g}

\providecommand{\e}{{\rm e}}
\providecommand{\comment}[1]{}

\newcommand{\beqn}[1]{\begin{eqnarray}\label{#1}}
\newcommand{\eeqn}{\end{eqnarray}}
\newcommand{\beq}[1]{\begin{equation}\label{#1}}
\newcommand{\eeq}{\end{equation}}

\makeatletter
\newcommand{\vast}{\bBigg@{4}}
\newcommand{\Vast}{\bBigg@{4.9}}
\makeatother

\providecommand{\LQGcost}{J}
\providecommand{\oLQGcost}{\bar{J}}
\providecommand{\CostXs}{q}
\providecommand{\CostUs}{r}
\providecommand{\Rate}{R}
\providecommand{\E}[1]{\mathbb{E}\left[ #1 \right]}

\providecommand{\fw}{f_W}
\providecommand{\Quantizer}{\mathcal{Q}}

\providecommand{\oD}{\bar{D}}
\providecommand{\oR}{\bar{R}}

\usepackage{pgf}
\providecommand*{\mynum}[1]{%
    \pgfmathprintnumber[
        fixed,
        precision=4,
        fixed zerofill=true,
        ]{#1}
}%

\providecommand{\PACKET}{\ell}
\providecommand{\packet}{l}

\providecommand{\E}[1]{\mathbb{E}\left[ #1 \right]}
\providecommand{\CE}[2]{\mathbb{E}\left[ #1 \middle| #2 \right]}

\providecommand{\e}{\text{e}}

\providecommand{\Interval}{\mathcal{I}}
\providecommand{\interval}[1]{\left[ 1 : #1 \right]}
\providecommand{\INTerval}[2]{\left[ #1 : #2 \right]}
\newcommand{\mT}{\interval{T}}
\providecommand{\VarW}{\sigma_W^2}

\usepackage{graphicx}
\usepackage{ifthen}

\usepackage{mathtools}
\mathtoolsset{showonlyrefs=true}

\providecommand{\ColumnNum}{1}
\newcommand{\col}{\ifthenelse{\equal{\ColumnNum}{1}}}

\begin{document}

\title{Algorithms for Optimal Control with \\ Fixed-Rate Feedback}
\author{Anatoly Khina, Yorie Nakahira, Yu Su, Hikmet Y\i ld\i z, and Babak Hassibi
	\thanks{This work was done, in part, while A.~Khina was visiting the Simons Institute for the Theory of Computing. 
    This work has received funding from the European Union's Horizon 2020 research and innovation programme under the Marie Sk\l odowska-Curie grant agreement No 708932.
    The work of Y.~Nakahira was funded by grants from AFOSR and NSF, and gifts from Cisco, Huawei, and Google.
    The work of Y.~Su was supported in part by NSF through AitF-1637598.
	The work of B.~Hassibi was supported in part by the National Science Foundation under Grant CNS-0932428, Grant CCF-1018927, Grant CCF-1423663, and Grant CCF-1409204; in part by a grant from Qualcomm Inc.; in part by NASA's Jet Propulsion Laboratory through the President and Director's Fund; and in part by King Abdullah University of Science and Technology. 
    The material in this paper was
presented in part at the IEEE Conference on Decision and Control, Melbourne, VIC, Australia, Dec., 2017.}
    \thanks{A.~Khina was with the Department of Electrical Engineering, California Institute of Technology, Pasadena, CA~91125, USA. 
    He is now with the Department of Electrical Engineering-Systems, Tel Aviv University, Tel Aviv 6997801, Israel (e-mail: \texttt{anatolyk@eng.tau.ac.il}).}
    \thanks{Y.~Nakahira and Y.~Su are with the Department of Computing and Mathematical Sciences, California Institute of Technology, Pasadena, CA~91125, USA (\mbox{e-mails}: \texttt{\{ynakahir,suyu\}@caltech.edu}).}
    \thanks{H.~Y\i ld\i z and B.~Hassibi are with the Department of Electrical Engineering, California Institute of Technology, Pasadena, CA~91125, USA (e-mails: \texttt{\{hyildiz,hassibi\}@caltech.edu}).}
} 

\maketitle


\begin{abstract}
We consider a discrete-time linear quadratic Gaussian networked control setting where the (full information) observer and controller are separated by a fixed-rate noiseless channel. The minimal rate required to stabilize such a system has been well studied. However, for a given fixed rate, how to quantize the states so as to optimize performance is an open question of great theoretical and practical significance. We concentrate on minimizing the control cost for first-order scalar systems. To that end, we use the Lloyd--Max algorithm and leverage properties of logarithmically-concave functions and sequential Bayesian filtering to construct the optimal quantizer that greedily minimizes the cost at every time instant. By connecting the globally optimal scheme to the problem of scalar successive refinement, we argue that its gain over the proposed greedy algorithm is negligible. This is significant since the globally optimal scheme is often computationally intractable. All the results are proven for the more general case of disturbances with logarithmically-concave distributions and rate-limited time-varying noiseless channels. We further extend the framework to event-triggered control by allowing to convey information via an additional ``silent symbol'', \ie, by avoiding transmitting bits; by constraining the minimal probability of silence we attain a tradeoff between the transmission rate and the control cost for rates below one bit per sample.
\end{abstract}

\allowdisplaybreaks

\section{Introduction}
\label{s:intro}

The demand for new and improved control techniques over unreliable communication links is constantly growing, due to the rise of emerging opportunities in the Internet of Things realm, as well as due to new and surprising applications in Biology and Neuroscience.

\begin{figure}
    \center
    \newcommand{\state}{X_t}
    \newcommand{\driveNoise}{W_t}
    \newcommand{\channelIn}{\PACKET_t}
    \newcommand{\contAction}{U_t} 
    \newcommand{\plant}{X_{t+1} = a \state + \driveNoise + \contAction}
    \newcommand{\rate}{\Rate_t \text{ bits}}
    \resizebox{\columnwidth}{!}{\begin{tikzpicture}[auto, arrow/.style={very thick, ->, >=stealth'},node distance=.2\columnwidth,>=latex']
    \node [coord] (input) {};
    \node [plant, right of = input, node distance=.36\columnwidth] (plant) 
    {$\begin{array}{c} 
    \text{Plant} 
    \\ \plant \end{array}$
    };
    \node [coord, right of=plant, node distance = .45\columnwidth] (sum) {};
    \node [coord, right of=sum, node distance = .2\columnwidth] (midoutput) {};
    \node [coord, right of=midoutput, node distance = .12\columnwidth] (output) {};

    \node [block, below of=sum, node distance = .2\columnwidth] (enc) {Observer};
    \node [block, below of=plant] (dec) {Controller};

    \draw[arrow] (sum) -- node (yArrow) {$\state$} (output);
    \draw[-,very thick] (plant) -- node {} (sum);

    \draw[arrow] (input) -- node {$\driveNoise$} (plant);
    \draw[arrow] (midoutput) |- node[above] {} (enc);
    \draw[arrow] (dec) -- node {$\contAction$} (plant);
    \draw[BitPipe] (enc) -- node [above] {$\channelIn$} node [below] {$\rate$} (dec);
    
\end{tikzpicture}
    }
    \caption{A scalar linear stochastic NCS with a communication channel of $R_t$ bits at time $t$.}
    \label{fig:system}
\end{figure}
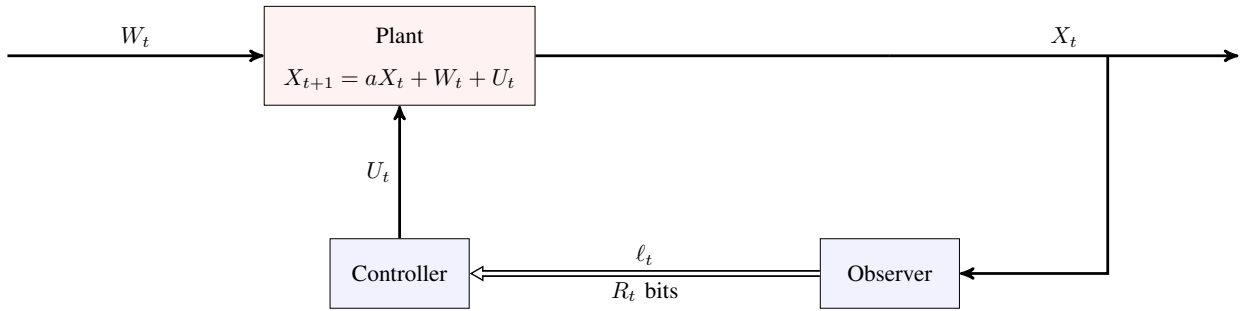

One of the most widely studied such \emph{networked control system} (NCS) setups is that of control over discretized packeted communication channels~\cite{NetworkedControlSurvey_ProcIEEE,SchenatoSinopoliFranceschettiPoolaSSS,GuptaDanaHespanhaMurrayHassibi_EstimationControl,Nair:L1,YukselBasarBook}. This setup can be further divided into two regimes: 
fixed-rate feedback---where exactly $\Rate$ bits can be noiselessly conveyed from the observer/encoder to the controller/decoder~\cite{TatikondaMitterQuantization,Yuksel:FixedRateControl} (see \figref{fig:system}), 
and variable-rate feedback---where $\Rate$ bits are available \emph{on average} and the observer/encoder can decide how many bits to allocate at each time instant~\cite{NairEvans:Quantization}.

For each of these scenarios, additional information can be conveyed through \textit{event-triggering} by allowing to remain \textit{silent}, 
\ie, not to send any information; see, \eg, 
\cite{KofmanBraslavsky:CDC2006,khojasteh:Allerton2016,khojasteh:CDC2017,EventTriggered:PearsonHespanhaLiberzon} 
and the references therein.

Although much effort has been put into determining the conditions for the stabilizability of such systems, less so has been done for determining the optimal attainable control costs---which are of great importance in practice---with several notable exceptions~\cite{TatikondaSahaiMitter,SilvaDerpichOstergaard:ECDQ4Control,KostinaHassibi:Allerton2016}.

Other effects that are encountered in practice when using packet-based protocols are those of \emph{packet erasures} (or \emph{packet drops}) and \emph{delayed packet arrivals}.
Consequently, much attention has been devoted to studying the impact these effects have on the performance of networked systems in an idealized setup where the quantization rate is {\em infinite}~\cite{Sinopoli:IntermittentObservations,GuptaDanaHespanhaMurrayHassibi_EstimationControl,SchenatoSinopoliFranceschettiPoolaSSS}.

A noteworthy effort to treat the case of finite-rate packets with packet drops was made by  Minero~\etal~\cite{MineroFranceschettiDeyNair}. To that end, they considered an even more general case where a time-varying rate ``budget'' $R_t$ (see \figref{fig:system}) is provided at every time step, and is determined and revealed just before transmission; 
a packet erasure corresponds to a zero-rate budget, implying that this scenario encompasses the packet-erasure setting.

In this work, we construct algorithms for the setting of time-varying feedback rate budget, presented in \secref{s:model}, along with its important special case of fixed-rate feedback.

However, in contrast to the works of Minero \etal~\cite{MineroFranceschettiDeyNair} and  Y\"uksel~\cite{Yuksel:FixedRateControl}, which concentrated on the conditions for system stabilizability using \textit{adaptive} uniform and logarithmic quantizers,\footnote{It is impossible to stabilize an unstable system using fixed-rate static quantization if the distributions of the disturbances or the initial state have unbounded supports~\cite[Sec.~III-A]{Nair:L1}.} respectively, 
we attempt to optimize the control~cost.

To that end, we concentrate our attention in \secref{s:greedy} 
on the class of disturbances that have logarithmically-concave (log-concave) probability density functions (PDFs) (the Gaussian PDF being an important special case), 
for which the Lloyd--Max algorithm~\cite[Ch.~6]{GershoGrayBook} 
is known to converge to the optimal quantizer~\cite{TrushkinConditions:Fleischer,TrushkinConditions:Kieffer,TrushkinConditions:Trushkin}.\footnote{Assuming contiguous cells; see~\remref{rem:MRSQ:regular}.\label{foot:contiguous-cells}}
Using Lloyd--Max quantization at every step, proposed previously by Bao \etal~\cite{LloydMax4Control:BaoSkoglundJohansson} and by Nakahira~\cite{Nakahira:L1LQG} (albeit without any optimality claims), 
and proving (\textit{a la} sequential Bayesian filtering~\cite{BayesianFilterBook}) that the resulting system state---which is composed of the scaled sums of quantization errors of the previous steps and the new disturbances---continues to have a log-concave PDF, leads us to an optimal greedy algorithm.

To support rates below one bit per sample, we extend the algorithm to the event-triggered control scenario in \secref{s:event-trig}. 
By adding another cell that corresponds to ``silence'' and constraining the probability 
of this cell to a minimal value, 
we are able to control the average rate of the scheme (which is equal to the sum of the probabilities of the remaining cells).

To tackle the more challenging task of designing a globally optimal quantizer, 
we recast the problem as that of designing an optimal quantizer for the problem of \emph{sequential coding of correlated  sources}~\cite{ViswanathanBerger:Streaming:IT} (see also \cite{StreamingWithFB:ITW2017,StreamingWithFB:CNS} and references therein).

An extreme variant of this problem is provided by that of linear quadratic regulator (LQR) control, 
in which the only randomness in the system happens in the intial state (which is again assumed to have log-concave PDF). 
We show in \secref{s:LQR} that this problem is equivalent to that of \emph{successive refinement}~\cite{EquitzCover91}, 
which can be regarded as a special case of sequential coding of correlated sources.

Surprisingly, for the latter, a computationally plausible variant of the Lloyd--Max algorithm exists~\cite{GeneralizedLloydMax:SR} that is known to achieve globally optimal performance for log-concave functions~\cite{DumitrescuWu:ScalarSuccessiveRefinement}.\footref{foot:contiguous-cells}
Furthermore, using the classical Bennett approximated quantization law~{\cite[Ch.~6.3]{GershoGrayBook}, \cite{GrayNeuhoff:Quantization:IT1998}}, 
we argue, in \secref{ss:Bennett}, that in the limit of high rates, 
the greedy algorithm is in fact optimal.

Although greedy optimization was demonstrated to be suboptimal~\cite{Fu:GreedySuboptimal} (outside of the high-rate regime), 
simulations for the LQR case show that the gain of the globally optimal algorithm over the optimal greedy one is modest even at low rates (for which the gain is expected to be the largest). This, in turn, suggests that the optimal greedy algorithm will remain close in performance to the optimum for the more general case where the state is driven by i.i.d. log-concave disturbances, which includes linear quadratic Gaussian (LQG) control.

We provide numerical performance evaluations in \secref{s:numeric}, 
and conclude the paper in \secref{s:discuss}.


\section{Problem Setup}
\label{s:model}

In this work, we consider the control--communication setup depicted in \figref{fig:system}. 
We use a discrete-time
model spanning the time interval $\mT \triangleq \{ 1, 2, \cdots, T\}$ for $T \in \nats$, 
where $\INTerval{i}{j} \triangleq \{i, i+1, \ldots, j\}$ for $i, j \in \ints$, 
such that $i \leq j$.
The plant is a discrete-time linear scalar stochastic system 
\begin{align}
\label{eq:plant}
	X_{t+1} &= a X_t + W_t + U_t ,
 	& t \in \INTerval{0}{T-1},
\end{align}
where $X_t, W_t, U_t \in \reals$ are the system state, disturbance and control action at time $t$, respectively.
We consider two setups for the disturbance sequence $\{ W_t \}$:

\begin{itemize}
\item
	\textbf{Independent and identically distributed (i.i.d.):}
    $\{W_t\}$ are i.i.d. according to a known log-concave PDF $\fw(w)$.
\item
	\textbf{LQR:}
    $W_0$ is distributed according to a known log-concave PDF $\fw(w)$; 
    $W_t = 0$ for all $t > 0$.
\end{itemize}
We further denote the variance of $\fw$ by $\VarW$ and assume, w.l.o.g., that it has zero mean.

\begin{defn}[Log-concave function; see~\cite{ConvexityBook}]
\label{def:log-concave}
	A function $f : \reals \to \reals_{\geq 0}$ is said to be log-concave
    if its logarithm $\log \circ f$ is concave:
    \begin{align}
    \label{eq:log-concave}
    	\log f \big( \lambda x + (1 - \lambda) y \big) \geq \lambda \log f(x) + (1 - \lambda) \log f(y) ,
    \end{align}
    for all $\lambda \in [0, 1]$ and $x, y \in \reals$;
    we use the extended definition that allows $f(x)$ to assign zero values, \ie,
    $\log f(x) \in \reals \cup \{-\infty\}$ is an extended real-value function that can take the value $-\infty$.
\end{defn}

\begin{remark}
	The Gaussian PDF is a log-concave function over $\reals$ and constitutes an important special case.
\end{remark}

We assume the observer has perfect access to $x_t$ at time $t$. 
However, in contrast to classical control settings, the observer is not co-located with the controller and communicates with it instead via a noiseless channel of \textit{data rate} $\Rate_t$. That is, at each time~$t$, the observer, which also takes the role of the encoder $\mE_t$, can perfectly convey a message (or ``index'') of $\Rate_t$ bits, $\PACKET_t \in \INTerval{0}{2^{\Rate_t}-1 }$, of the past states, to the controller:
\begin{align}
\label{encoder}
	\PACKET_t =\mE_t (  X^t ) , 
\end{align}
where we denote $a^t \triangleq (a_1, a_2, \ldots, a_t)$ and use the convention that $a^t = \emptyset$ for $t \leq 0$.
We further set $\PACKET_0  = \PACKET_T = 0$.

The controller at time $t$, which also takes the role of the decoder $\mD_t$, recovers the observed codeword $\PACKET_t$ and uses it to generate the control action
\begin{align}
\label{decoder}
	U_t = \mD_t \left( \PACKET^t  \right).
\end{align} 

The exact value of $\Rate_t$ is revealed to the encoder prior to the computation of $\ell_t$ and is inferred by the decoder upon receiving $\ell_t$. The statistics of $\Rate_t$ impact system performance but do not affect the 
greedy optimality guarantees of the proposed algorithm of \secref{s:greedy}.

\begin{remark}[Packet-erasure channel]
\label{rem:packet-erasure}
A packet-erasure can be modeled by $\Rate_t = 0$. Hence, the time-varying data rate model subsumes the packet-erasure scenario~\cite{MineroFranceschettiDeyNair}. 
\end{remark}

Our goal is to minimize the following average-stage linear quadratic (LQ) cost upon reaching the time horizon~$T \in \nats$: 
\begin{subequations}
\label{cost}
\noeqref{cost:explicit,cost:J}
\begin{align}
	\oLQGcost_T &\triangleq
    \frac{1}{T}\E{\CostXs_T X^2_T +  \sum_{t = 1}^{T-1} \left( \CostXs_t X^2_t + \CostUs_t U^2_t \right)}
\label{cost:explicit}
\\ &= \frac{1}{T} \sum_{t = 1}^T \LQGcost_t \,,
\label{cost:J}
\end{align}
\end{subequations}
where $\{\LQGcost_t\}$ are the instantaneous costs 
\begin{subequations}
\label{eq:cost:instant}
\noeqref{eq:cost:instant:mid,eq:cost:instant:last}
\begin{align}
	\LQGcost_t &\triangleq \E{\CostXs_t X^2_t + \CostUs_t U^2_t} , &t\in \interval{T-1} ,
\label{eq:cost:instant:mid}
 \\ \LQGcost_T &\triangleq \E{\CostXs_T X^2_T}.
\label{eq:cost:instant:last}
\end{align}
\end{subequations}
The weights $\{\CostXs_t\}$ and $\{\CostUs_t\}$ penalize the state deviation and actuation effort, respectively. 


\section{Optimal Greedy Control}
\label{s:greedy}

In this section we consider the i.i.d.\ disturbance setting.
We recall the Lloyd--Max algorithm and its optimality guarantees in~\secref{ss:LM:quantizer}, which are subsequently used in \secref{ss:LM:control}
to construct a greedy optimal control policy.

\subsection{Quantizer Design}
\label{ss:LM:quantizer}

\begin{defn}[Scalar quantizer]
\label{def:quantizer}
    A scalar quantizer $\Quantizer$ of rate $\Rate$ is described by 
    an encoder~$\mE_\Quantizer: \reals \to \INTerval{0}{2^\Rate-1}$ 
    and a decoder~\mbox{$\mD_\Quantizer : \INTerval{0}{2^\Rate-1} \to c \triangleq \{c[0], \ldots, c[2^\Rate-1]\} \subset \reals$}.
    We define the quantization operation $\Quantizer: \reals \rightarrow c$ as the composition of the encoding and decoding operations: 
    $\Quantizer = \mD_\Quantizer \circ \mE_\Quantizer$.\footnote{The encoder and decoder that give rise to the same parameter are unique up to a permutation of the labeling of the index $\ell$.}
    The \textit{reproduction~points} $\{c[0], \ldots, c[2^\Rate-1]\}$ are assumed 
    to be ordered, without loss of generality:\footnote{If some inequalities are not strict, 
    then the quantizer can be reduced to a lower-rate quantizer.}
    \begin{align}
        c[0] < c[1] < \cdots < c[2^\Rate-1] .
    \end{align}
    We denote by $\cI[\ell]$ the collection of all points that are mapped to index $\ell$ (equivalently to the reproduction point $c[\ell]$): 
    \begin{align}
    	\cI[\ell] &\triangleq \{ x | x \in \reals, \mE_\Quantizer = \ell \}
	 \\ &= \{ x | x \in \reals, \Quantizer = c[\ell] \} .
    \end{align}
\end{defn}

	We shall concentrate on the class of \textit{regular quantizers}, defined next.
\begin{defn}[Regular quantizer]
\label{def:quantizer:regular}
	A scalar quantizer is \emph{regular} if every cell $\cI[\ell], \ell \in \INTerval{0}{2^\Rate-1}$, is a contiguous interval that contains its reproduction point $c[\ell]$:
    \begin{align}
    	c[\ell] \in \cI[\ell] &= \left[ p[\ell], p[\ell + 1] \right) , & \ell \in \INTerval{0}{2^\Rate - 1},
    \end{align}
    where $p \triangleq \left\{ p[0], \ldots, p[2^\Rate] \right\}$ is the set of \textit{partition levels}---the boundaries of the cells.
    Hence, a regular scalar quantizer can be represented by the input partition-level set $p$ and the reproduction-point set $c \triangleq \left\{ c[0], \ldots, c[2^\Rate-1] \right\}$.
    We further take $p[0]$ and $p[2^\Rate]$ to be the left-most and right-most values of the support of the source's PDF.
\end{defn}

The cost we wish to minimize is the mean squared error distortion between the source $W$ with a given PDF $\fw$ and its quantization $\Quantizer(W)$:
\begin{subequations}
\label{eq:dist-measure}
\noeqref{eq:dist-measure:E,eq:dist-measure:regular}
\begin{align}
	D &\triangleq \E{ \left\{ W - \Quantizer(W) \right\}^2 }
\label{eq:dist-measure:E}
 \\ &= \sum_{\ell = 0}^{2^\Rate - 1} \int_{p[\ell]}^{p[\ell+1]} \left( w - c[\ell] \right)^2 \fw(w) dw .
\label{eq:dist-measure:regular}
\end{align}
\end{subequations}
Denote by $D^*$ the minimal achievable distortion $D$; 
the \textit{optimal quantizer} is the one that achieves $D^*$.

\begin{remark}
\label{rem:continuous:PDF}
	We shall concentrate on log-concave PDFs $\fw$, which are therefore continuous~\cite{ConvexityBook}.
    Hence, the inclusion or exclusion of the boundary points in each cell does not affect the distortion of the quantizer, meaning that the boundary points can be broken systematically.
\end{remark}

\begin{remark}
	If the input PDF has an infinite/semi-infinite support, then the leftmost and/or rightmost intervals of the quantizer are open ($p[0]$ and/or $p[2^\Rate]$ take infinite values).
\end{remark}

The optimal quantizer satisfies the following necessary conditions~\cite[Ch.~6.2]{GershoGrayBook}.

\begin{prop}[Centroid condition]
\label{prop:Centroid}
	For a fixed partition-level set $p$ (fixed encoder), 
    the reproduction-point set $c$ (decoder) that minimizes the distortion $D$~\eqref{eq:dist-measure} is 
	\begin{align}
	\label{eq:3_opt_c}
		c[\ell] &= \E{w \, \big| \, p[\ell] < w  \leq p[\ell+1]} , & \ell \in \INTerval{0}{2^\Rate-1}. \ \ \ 
	\end{align}
\end{prop}

\begin{prop}[Nearest neighbor condition]
\label{prop:NN}
	For a fixed reproduction-point set $c$ (fixed decoder), 
    the partition-level set $p$ (encoder) that minimize the distortion $D$~\eqref{eq:dist-measure} is
	\begin{align}
	\label{eq:3_opt_p}
		p[\ell] &= \frac{c[\ell-1] + c[\ell]}{2} , & \ell \in \interval{2^\Rate-1},
	\end{align}
  where the leftmost/rightmost boundary points $p[0]/p[2^\Rate]$ are equal to the smallest/largest values of the support of $\fw$. 
\end{prop}

The optimal quantizer must simultaneously satisfy both \eqref{eq:3_opt_c} and \eqref{eq:3_opt_p};
iterating between these two necessary conditions gives rise to the Lloyd--Max algorithm.

\begin{algo}[Lloyd--Max quantization]
\label{algo:LM}
\ 

  \textbf{\textit{Initial step}.}
  Pick an initial partition-level set $p$. 

  \textbf{\textit{Iterative step}.} Repeat the two steps 
    \begin{enumerate}
    \item 
    	Fix $p$ and set $c$ as in \eqref{eq:3_opt_c}, 
    \item 
    	Fix $c$ and set $p$ as in \eqref{eq:3_opt_p}, 
    \end{enumerate}
    interchangeably, until the decrease in the distortion $D$ per iteration goes below a desired accuracy threshold.
\end{algo}

Props.~\ref{prop:Centroid} and \ref{prop:NN} suggest that the distortion at every iteration decreases;
since the distortion is bounded from below by zero, the Lloyd--Max algorithm is guaranteed to converge to a local optimum.

Unfortunately, multiple local optima may exist in general (\eg, Gaussian mixtures with well separated components), rendering the algorithm sensitive to the initial choice~$p$.

Nonetheless, sufficient conditions for the existence of a unique global optimum were established in~\cite{TrushkinConditions:Fleischer,TrushkinConditions:Kieffer,TrushkinConditions:Trushkin}. 
These guarantee that the algorithm converges to the global optimum for any initial choice of $p$. 
An important class of PDFs that satisfy these conditions is that of log-concave PDFs.

\begin{thm}[\!\! \cite{TrushkinConditions:Fleischer,TrushkinConditions:Kieffer,TrushkinConditions:Trushkin}]
\label{lem:LM_convergence}
	Let the source PDF $\fw$ be log-concave. 
    Then, the Lloyd--Max algorithm converges to a unique solution that minimizes the mean squared error distortion~\eqref{eq:dist-measure}.
\end{thm}

\subsection{Controller Design}
\label{ss:LM:control}

We now describe the optimal greedy control policy. 
To that end, we make use of the following lemma that extends the separation principle of estimation and control to networked control.

\begin{lemma}[Control--estimation separation~{\!\!}{\cite{QuantizedLQG:Fischer82}, \cite{TatikondaSahaiMitter}}]
\label{lem:LQG:separation}
	Consider the general cost problem~\eqref{cost} with independent disturbance elements $\{w_t\}$ of variances $\{\sigma_{W_t}^2\}$. 
    Then, the optimal controller 
    has the structure
    \begin{align}
    \label{eq:lem:LQG:separation:controller}
    	U_t &= -k_t  \hX_t ,
    \end{align}
    where 
    \begin{align}
		k_t &= \frac{s_{t+1}}{s_{t+1} + \CostUs_t}  a
    \end{align}
    is the optimal LQR control gain and $\hX_t \triangleq \CE{X_t}{\PACKET^t}$, 
    and $s_t$ satisfies the dynamic Riccati backward recursion~\cite{BertsekasControlVol1}:
    \begin{align}
    \label{eq:lem:LQG:Lt}
    	s_t &= \CostXs_t + \frac{s_{t+1} \CostUs_t}{ s_{t+1} + \CostUs_t} a^2  ,
    \end{align}
    with $s_T = \CostXs_T$ and $s_{T+1} = k_T = 0$. 
	Moreover, this controller achieves a cost of\footnote{Recall that $\Rate_T = 0$ and $\PACKET_T = 0$ for the definition of $\hx_T$, as no transmission or control action are performed at time $T$.}
    \begin{align}
    \label{eq:LQR:cost}
     	\oLQGcost_T  = \frac{1}{T} \sum_{t = 1}^{T} \Big( s_t \sigma_{W_t}^2 + g_t \E{(X_t  -\hX_t )^2} \Big) , 
    \end{align}
    with 
    \begin{align}
    \label{eq:G_k_def}
        g_t =  s_{t+1} a^2 - s_t + \CostXs_t \,.
    \end{align}
\end{lemma}

\begin{remark}
	\lemref{lem:LQG:separation} holds true for any memoryless channel, with $\hX_t = \CE{X_t}{\PACKET^t}$, 
    where $\PACKET_t$ is the channel output at time~$t$.
\end{remark}

The optimal greedy algorithm minimizes the estimation distortion $\E{(X_t  -\hX_t)^2}$ at time $t$, 
without regard to its effect on future distortions. 
To that end, at time $t$, 
the encoder and the decoder calculate the PDF of $x_t$ conditioned on $\PACKET^{t-1}$, $f_{X_t | \PACKET^{t-1}}$ via sequential Bayesian filtering~\cite{BayesianFilterBook}, and apply the Lloyd--Max quantizer to this PDF. We refer to $f_{X_t | \PACKET^{t-1}}$ and to $f_{X_t | \PACKET^t}$ as the \textit{prior} and \text{posterior PDFs}, respectively.

\begin{algo}[Optimal greedy control]
\label{algo:greedy}
\
    
   \vspace{.3\baselineskip}
    \textbf{\textit{Initialization}.}
    Both the encoder and the decoder set 
    \begin{enumerate}\addtolength{\itemsep}{.15\baselineskip}
    \item
    \label{itm:greedy-algo:coeffs}
    	$\{s_t, k_t | t \in \mT \}$ as in \lemref{lem:LQG:separation}, 
    	for the given $T$, $\{\CostXs_t\}$, $\{\CostUs_t\}$ and $a$.
    \item
    	$\PACKET_0 = X_0 = U_0 = 0$.
    \item
    \label{itm:greedy-algo:prior}
    	The prior PDF: 
    	$f_{X_1| \PACKET_0}(x_1 | 0) \equiv \fw(x_1)$.
    \end{enumerate}
   
	\vspace{.3\baselineskip}
    \textbf{\textit{Observer/Encoder}.}
    At time $t \in \interval{T-1}$:
    \begin{enumerate}
    \item
    	Observes the current state $x_t$.
    \item
    \label{itm:algo:greedy:ENC:LM}
    	Runs the Lloyd--Max algorithm (\algoref{algo:LM}) with respect to the prior PDF $f_{X_t | \PACKET^{t-1}}$ to obtain the quantizer $\Quantizer_t(x_t)$ of rate $\Rate$; denote its partition and reproduction sets by $p_t$ and $c_t$, respectively, 
        and the cell corresponding to $p_t[\packet]$---by $\Interval_t[\packet]$.
	\item
    \label{itm:algo:greedy:quantize}
    	Quantizes the system state $x_t$ [recall \defnref{def:quantizer}]: 
        \begin{subequations}
        \label{eq:greedy:enc}
        \noeqref{eq:greedy:enc:dec}
        \begin{align}
        	\packet_t &= \mE_{\Quantizer_t}(x_t) =: \mE_t(x^t), 
        \label{eq:greedy:enc:enc}
		 \\ \hx_t &= \Quantizer_t(x_t) =
        \mD_{\Quantizer_t}(\packet_t),
        \label{eq:greedy:enc:dec}
        \end{align}
        \end{subequations}
        where $\mE_t(x^t)$ is the overall action of the observer/encoder at time $t$ as defined in~\eqref{encoder}.
	\item
    	Transmits the quantization index $\packet_t$.
	\item
    \label{itm:algo:greedy:ENC:posterior}
    	Calculates the posterior PDF $f_{X_t | \PACKET^t}(x_t | \packet^t)$:
        \begin{align}
		\!\!\!\!\!\!\!\!\!
        	f_{X_t | \PACKET^t}(x_t | \packet^t) &=
            \begin{cases}
            	f_{X_t|\PACKET^{t-1}}(x_t | \packet^{t-1}) / \gamma , & x_t \in 
                \Interval_t[\packet_t] ,
			 \\ 0 & \text{otherwise} , \ \ \ 
            \end{cases}
        \label{eq:posterior}
        \end{align}
        where\footnote{We use here the regularity assumption.}
        \begin{align}
			\gamma &\triangleq \int_{p_t[\packet_t]}^{p_t[\packet_t+1]} f_{X_t|\PACKET^{t-1}}(\alpha | \packet^{t-1}) d \alpha 
		\nonumber
		\end{align}
        is a normalization factor.
        
	\item
    \label{itm:algo:greedy:ENC:prior}
	\vspace{.3\baselineskip}
        Determines the prior PDF of time $t+1$ using \eqref{eq:plant} and the control action~\eqref{decoder} $u_t = - k_t \hx_t := \mD_t(\hx^t)$:
        \begin{align}
        \begin{aligned}
            &f_{X_{t+1} | \PACKET^t} \left( x_{t+1} | \packet^t \right) 
		 \\ &\quad = \frac{1}{|a|} f_{X_t | \PACKET^t} \left( \frac{x_{t+1} - u_t}{a} \middle| \packet^t \right)  * \fw \left( x_{t+1} \right) , \quad
        \end{aligned}
        \label{eq:prior_compute}
        \end{align}
        where `$*$' denotes the convolution operation, and the two convolved terms correspond to the PDFs of the quantization error $a (X_t - \hX_t)$ and the disturbance $W_t$.
    \end{enumerate}

   \vspace{.3\baselineskip}
	\textbf{\textit{Controller/Decoder}.}
    At time $t \in \interval{T-1}$:
    \begin{enumerate}
    \item
    	Runs the Lloyd--Max algorithm (\algoref{algo:LM}) with respect to the prior PDF $f_{X_t | \PACKET^{t-1}}$ 
        as in Step \ref{itm:algo:greedy:ENC:LM} of the observer/encoder protocol.
    \item
    	Receives the index $\packet_t$.
	\item
    	\label{itm:algo:greedy:DEC:hX}
    	Reconstructs the quantized value: $\hx_t = \mD_{\Quantizer_t}(\packet_t)$.
	\item
    	\label{itm:algo:greedy:DEC:U}
    	Generates the control actuation 
        \begin{align}
        \label{eq:greedy:dec}
        	u_t = - k_t \hx_t := \mD_t(\hx^t).
        \end{align}
    \item
    	Calculates the posterior PDF $f_{X_t | \PACKET^t}$ 
        and the next prior PDF $f_{X_{t+1} | \PACKET^t}$ 
        as in Steps \ref{itm:algo:greedy:ENC:posterior} and \ref{itm:algo:greedy:ENC:prior} 
        of the observer/encoder protocol.
    \end{enumerate}
\end{algo}

\begin{thm}
\label{thm:greedy_optimality}
	Let $\fw$ be a log-concave PDF (recall \defnref{def:log-concave}).
	Then, \algoref{algo:greedy} is the optimal greedy control policy.
\end{thm}

The following is an immediate consequence of the log-concavity of the Gaussian PDF.
\begin{corol}
	Let $\fw$ be a Gaussian PDF.
	Then, \algoref{algo:greedy} is the optimal greedy control policy.
\end{corol}

Recall that the Lloyd--Max Algorithm converges to the global minimum for log-concave PDFs. 
Consequently, in order to prove \thmref{thm:greedy_optimality}, 
it suffices to show that all the prior PDFs 
$\{f_{X_t | \PACKET^t} : t \in \mT \}$ are log-concave. 
This, in turn, relies on the following log-concavity properties.

\begin{assert}[Log-concave function properties~\cite{ConvexityBook}]
\label{assert:LogConcaveProps}
	Let $f(x)$ and $g(x)$ be log-concave functions over $\reals$.
    Then, the following are also log-concave functions:
    \begin{itemize}\addtolength{\itemsep}{.3\baselineskip}
    \item
    	\textbf{Affinity:} $c f(a x + b)$ for any constants $a, b, c \in \reals$.
    \item
    	\textbf{Truncation:} $\begin{cases} f(x) & x \in I, \\ 0 & \mathrm{otherwise}, \end{cases}$
        \quad for any interval $I$, possibly (semi-)infinite.
	\item
    	\textbf{Convolution:} $f(x) * g(x)$.
    \end{itemize}
\end{assert}

Now we are ready to prove \thmref{thm:greedy_optimality}.

\begin{IEEEproof}[Proof of \thmref{thm:greedy_optimality}]
	We use mathematical induction to show that both of the following conditions hold for any time $t \in \mT$: 
    \begin{enumerate}[(i)]
    \item
    \label{cnd:log-concave:prior}
    	The prior PDF $f_{X_t | \PACKET^{t-1}}(x_t | \packet^{t-1})$ is log-concave in $x_t$ for any realization $\packet^{t-1} \in \bigtimes\limits_{i=1}^{t-1} \INTerval{0}{2^{\Rate_i}-1}$.
    \item
    \label{cnd:log-concave:LM-opt}
        Given the past policies $\mE^t$ and  $\mD^t$ of \eqref{eq:greedy:enc:enc} and \eqref{eq:greedy:dec} of \algoref{algo:greedy}, 
        it minimizes the instantaneous cost~\eqref{eq:cost:instant} $\LQGcost_{t+1}$ at time $t+1$.
    \end{enumerate}

    \textit{Basic step ($t = 1$)}.
    From the initial condition $X_0 = 0$, 
    the optimal control action for $t=0$ is $U_0 = 0$, and hence $X_1 = W_0$.
     Since $\fw$ is log-concave from the model assumption, $X_1$ also has a log-concave PDF, 
     yielding \condref{cnd:log-concave:prior}.
    Consequently, 
    the quantizer $\Quantizer_1$ generated by the Lloyd--Max algorithm and the controller $U_1 = - a \hX_1 
	= -a \Quantizer_1(X_1)$ minimizes the instantaneous cost $\LQGcost_2$, yielding \condref{cnd:log-concave:LM-opt}.

    \textit{Inductive step}.
	Assuming Conds.~(i)-(ii) hold at time $t\in \nats$, we show below that they also hold at time $t+1$. 
    By the induction hypothesis, $f_{X_t | \PACKET^{t-1}}$ is log-concave. 
    Consequently, by~\thmref{lem:LM_convergence}, the Lloyd--Max Algorithm generates the quantizer $\Quantizer_t$ that 
    minimizes the cost
    $\LQGcost_{t+1}$. 
    This leads to \condref{cnd:log-concave:LM-opt}. 
   
   It only remains to show that \condref{cnd:log-concave:prior} holds. 
    Since log-concavity is preserved under truncation and affinity, 
    and $f_{X_t | \PACKET^{t-1}}$ is log-concave by the induction hypothesis, 
    the posterior PDF $f_{X_t | \PACKET^t}$ of \eqref{eq:posterior} is also log-concave for any realization of $\PACKET^t \in \bigtimes_{i=1}^t \INTerval{0}{2^{\Rate_i}-1}$; 
    this, along with the log-concavity of $\fw$ and the log-concavity preservation under affine transformations and convolution, guarantees the log-concavity of the next prior~\eqref{eq:prior_compute}, $f_{X_{t+1} | \PACKET^t}$, and completes the proof.
\end{IEEEproof}


\section{Event-triggered Control}
\label{s:event-trig}

In this section, we extend the greedy algorithm to the event-triggered control scenario.
Under this scenario, the encoder may either send a packet of a fixed predetermined rate or avoid transmission altogether. Avoiding transmission helps alleviating network congestion by conveying information ``by silence''.

We concentrate on the case of packets of a single bit, 
as in this regime the advantage of the algorithm is most pronounced and the exposition of the algorithm is the simplest. 
The two cells corresponding to the single-bit packet along with the silence symbol form a three-level algorithm.
We add a constraint $\delta$ on the minimal probability of the silent symbol; 
clearly, the average transmission rate is equal to $\oR \triangleq \E{\Rate_t} \equiv 1 - \delta$ in this case.
To minimize the average transmission rate, the silence symbol needs to  be assigned to the cell with the maximal probability:
\begin{align}
\label{eq:event-trig:bin-choice}
	\max_{\ell=0,1,2}\int_{p[\ell]}^{p[\ell+1]}\fw(w)dw \geq \delta,
\end{align}
where the cell-index $\ell$ that achieves the maximum in \eqref{eq:event-trig:bin-choice} 
corresponds to the \textit{silent cell}; we denote this index by $\ell^*$.

Hence, the standard Lloyd--Max quantizer of \algoref{algo:LM} in each time step 
should be replaced by the following algorithm, which first checks whether standard three-level Lloyd--Max quantization satisfies the constraint~\eqref{eq:event-trig:bin-choice} and, if not, runs the algorithm with the constraint \eqref{eq:event-trig:bin-choice} imposed on a different cell each time, and chooses the one that achieves minimal average distortion.
With the constraint imposed on a particular cell, the algorithm iterates between two steps: choosing the optimum $c$ for a fixed $p$ and choosing the optimum $p$ for a fixed $c$. The first step is the same as the standard Lloyd-Max quantizer. For the second step, 
the Karush--Khun--Tucker (KKT) conditions are employed~\cite[Ch.~5]{BoydBook}.

\begin{algo}[Min.\ cell-probability constrained quantization]
\label{algo:LM_silence}

	\indent
	\textit{Unconstrained algorithm.}
    Apply \algoref{algo:LM}. If the constraint~\eqref{eq:event-trig:bin-choice} is satisfied for the resulting quantizer, use this quantization law.
    Else, set $p[0]$ and $p[3]$ to the smallest and largest values of the support of $\fw$, 
    and run the following. 
	\begin{enumerate}
    \item[0)] $\ell^* = 0$.
    	\begin{enumerate}
        \item
            Set $p[1]$ such that 
            \begin{align}
                \int_{p[0]}^{p[1]} \fw(w) dw = \delta .
            \end{align}
        \item
        	Compute $c[0]$ as in \eqref{eq:3_opt_c}. 
        \item
            Run \algoref{algo:LM} for the remaining two cells (with $p[0], p[1], c[0]$ remain fixed), to determine $p[2]$ and $c[2]$.
		\item
        	Denote the resulting overall quantizer and distortion by $\Quantizer_0$ and $D_0$, respectively.
        \end{enumerate}
      
      \item $\ell^* = 1$.

		\textbf{\textit{Initial step}.} 
		Pick an initial partition-level set $p$.
		
		\textbf{\textit{Iterative step}.} Repeat the following steps
		\begin{enumerate}
        \item 
            Fix $p$ and set $c$ as in \eqref{eq:3_opt_c}, 
        \item 
            Fix $c$ and set $p$ as in \eqref{eq:3_opt_p},
        \item
            If $p$ does not satisfy the constraint~\eqref{eq:event-trig:bin-choice}, set $p$, in accordance with the KKT conditions, as the solution of 
            {{\mathtoolsset{showonlyrefs=false}
                \begin{subnumcases}{\label{eq:event-trig:midcell}}
                    \displaystyle \delta = \int_{p[1]}^{p[2]} \fw(w) dw
                \label{eq:event-trig:midcell:prob}
                 \\ p[2] = \frac{c[0]-c[1]}{c[2]-c[1]} p[1] + \frac{c[2]^2-c[0]^2}{2(c[2]-c[1])} \qquad
                \label{eq:event-trig:midcell:partition}
                \end{subnumcases}
            }}
        \item
			If no solution to \eqref{eq:event-trig:midcell} exists, 
            replace \eqref{eq:event-trig:midcell:partition} with the choice that gives the smaller distortion out of $p[1]=p[0]$ and $p[2]=p[3]$,
		\end{enumerate}
		until the decrease in the distortion $D$ per iteration is below a desired accuracy threshold. Denote the resulting quantizer and distortion by $\Quantizer_1$ and $D_1$, respectively.
        
        
    \item$\ell^* = 2$.
    	\begin{enumerate}
        \item
            Set $p[2]$ such that 
            \begin{align}
                \int_{p[2]}^{p[3]} \fw(w) dw = \delta .
            \end{align}
        \item
        	Compute $c[2]$ as in \eqref{eq:3_opt_c}. 
        \item
            Run \algoref{algo:LM} for the remaining two cells (with $p[2], p[3], c[2]$ remain fixed), to determine $p[1]$ and $c[1]$.
		\item
        	Denote the resulting overall quantizer and distortion by $\Quantizer_2$ and $D_2$, respectively.
        \end{enumerate}
      
        
   	\item 
    	Set the quantizer to $\Quantizer_i$, where $i^*=\arg\min_{i=0,1,2} D_i$.
	\end{enumerate}
\end{algo}

Replacing the Lloyd--Max quantizer of \algoref{algo:LM}
with the constrained variant of \algoref{algo:LM_silence}
gives rise to the following event-triggered variant of \algoref{algo:greedy}.

\begin{algo}[Greedy event-triggered control]
\ 

    \textbf{\textit{Initialization}.}
    Both the encoder and the decoder 
    \begin{enumerate}\addtolength{\itemsep}{.15\baselineskip}
    \item
    	Run steps \ref{itm:greedy-algo:coeffs}--\ref{itm:greedy-algo:prior} of the initialization of \algoref{algo:greedy}.
	\item
    	Set $\delta = 1 - \oR$.\footnote{Recall that we assume $\oR \in (0, 1]$.}
	\end{enumerate}

   \vspace{.3\baselineskip}
    \textbf{\textit{Observer/Encoder}.}
    At time $t \in \interval{T-1}$:
    \begin{enumerate}
    \item
    	Observes $x_t$.
    \item
    \label{itm:algo:event-trig:ENC:LM}
    	Runs \algoref{algo:LM_silence} with respect to the prior PDF $f_{X_t | \PACKET^{t-1}}$ and the maximal probability constraint $\delta$ to obtain the quantizer $\Quantizer_t$; denote its partition and reproduction sets by $p_t$ and $c_t$, respectively, the index of the silent cell---by $\ell^*_t$, 
		and the cell corresponding to $p_t[\ell]$---by~$\Interval_t[\ell]$.
	\item
    	Quantizes the system state $x_t$ as in Step~\ref{itm:algo:greedy:quantize} of the observer/encoder protocol of \algoref{algo:greedy}.
	\item
    	If $\packet_t \neq \packet^*_t$, transmits the index $\packet_t$;
    	otherwise, remains silent.
	\item
    	Calculates the posterior PDF $f_{X_t | \PACKET^t}$ and the next prior PDF $f_{X_{t+1} | \PACKET^t}$
        as in Steps~\ref{itm:algo:greedy:ENC:posterior} and \ref{itm:algo:greedy:ENC:prior} 
        of the observer/encoder protocol of \algoref{algo:greedy}, respectively.
    \end{enumerate}

   \vspace{.3\baselineskip}
	\textbf{\textit{Controller/Decoder}.}
    At time $t \in \interval{T-1}$:
    \begin{enumerate}
    \item
    	Runs \algoref{algo:LM_silence} with respect to the prior PDF $f_{X_t | \PACKET^{t-1}}$ 
        as in Step \ref{itm:algo:event-trig:ENC:LM} of the observer/encoder protocol.
    \item
    	Receives the index $\packet_t$: in case of silence, recovers $\packet_t = \packet^*_t$.
	\item
    	Reconstructs the quantized value: $\hx_t = \mD_{\Quantizer_t}(\packet_t)$.
	\item
    	Generates the control actuation $u_t = - k_t \hx_t$.
	\item
    	Calculates the posterior PDF $f_{X_t | \PACKET^t}$ and the next prior PDF $f_{X_{t+1} | \PACKET^t}$
        as in Steps~\ref{itm:algo:greedy:ENC:posterior} and \ref{itm:algo:greedy:ENC:prior} 
        of the observer/encoder protocol of \algoref{algo:greedy}, respectively.
    \end{enumerate}
\end{algo}


\section{Globally Optimal LQR Control}
\label{s:LQR}

In this section, we study the LQR control setting, namely, 
the case where $W_0$ has a log-concave PDF $\fw$ and $W_t = 0$ for all 
$t \in \interval{T-1}$. 
Clearly, this is equivalent to the case of a random initial condition $X_0$ and $W_t \equiv 0$ for all $t \in \INTerval{0}{T-1}$, 
and is therefore referred to as LQR control. 

We construct a globally optimal control policy in \secref{ss:LQR:controller} by connecting the problem to that of scalar successive refinement~\cite{GeneralizedLloydMax:SR,DumitrescuWu:ScalarSuccessiveRefinement},
which is formulated and reviewed in~\secref{ss:LQR:SR}.
The resulting quantizers are commonly referred to as \textit{multi-resolution scalar quantizers} (MRSQs).

\subsection{Successive Refinement}
\label{ss:LQR:SR}

A $T$-step MRSQ successively quantizes a single source sample $W \in \reals$ with PDF $\fw$
using a series of quantizers $\Quantizer^T$ of rates $\Rate^T$: 
At stage $t \in \mT$, 
$\Rate_t$ bits are available 
for the re-quantization of the source $W$, 
and are encoded into an index $\PACKET_t \in \INTerval{0}{2^{\Rate_t}-1}$. 
$\PACKET_t$, along with all previous indices $\PACKET^{t-1}$, is then used for the construction of a refined description $\hW_t = \Quantizer_t(W)$.

\begin{defn}[MRSQ]
	A $T$-step MRSQ of rates $\Rate^T$ is described by a series of $T$ encoders $(\mE_{\Quantizer_1}, \ldots, \mE_{\Quantizer_T})$ 
    and a series of $T$ decoders $(\mD_{\Quantizer_1}, \ldots, \mD_{\Quantizer_T})$, 
    with $\mE_{\Quantizer_t} : \reals \to \INTerval{0}{2^{\Rate_t} - 1}$ and 
    $\mD_{\Quantizer_t} : \bigtimes_{i=1}^t \INTerval{0}{2^{\Rate_i} - 1} \to \{ c_t[0], c_t[1], \ldots, c_t[2^{\sum_{i=1}^t \Rate_i} - 1] \}$ serving as the encoder and decoder at time $t$, respectively.
    We define the quantization operation 
    $\Quantizer_t : \reals \to \{ c_t[0], c_t[1], \ldots, c_t[2^{\sum_{i=1}^t \Rate_i} - 1] \}$, 
    at time step $t$,
    as the composition of all the encodings until time $t$ with the decoding at time $t$:
    $\Quantizer_t = \mD_{\Quantizer_t} \circ \left( \mE_{\Quantizer_1}, \ldots, \mE_{\Quantizer_t} \right)$.
\end{defn}

This definition means that, although the overall effective rate of the quantizer at time $t$ is $\sum_{i=1}^t \Rate_i$, 
only the last $\Rate_t$ bits, corresponding to $\PACKET_t$, are determined during time step $t$. At the decoder, these bits are appended to the previously determined and received $\sum_{i=1}^{t-1} \Rate_i$ bits (corresponding to $\PACKET^{t-1}$), for the construction of a description of $W$ at time $t$, $\hW_t = \Quantizer_t(W)$.

\begin{defn}[Regular MRSQ]
\label{def:convexMRSQ}
	A $T$-step MRSQ is \textit{regular} if the quantizer at each step $t \in \mT$ is regular 
    and the partitions of subsequent stages are nested, as follows.
    For each time $t \in \{2, \ldots, T\}$:
    \begin{align}
    \label{eq:def:MRSQ:p}
    	p_t \left[ \ell \cdot 2^{\ver{\Rate_t}{\Rate}} \right] &= p_{t-1}[\ell] , & 
         \ell \in \INTerval{0}{\left( \sum_{i=1}^{t-1} \Rate_i \right) - 1} ,
         \quad
    \end{align}
    where $p_t$ is the partition-level set of the quantizer at time $t$.
\end{defn}   

\begin{remark}
\label{rem:MRSQ:partitions}
	The relation in \eqref{eq:def:MRSQ:p} implies that, given $p_T$, 
    the partitions of all the previous stages can be deduced.
\end{remark}

\begin{remark}[Optimality of regular MRSQs]
\label{rem:MRSQ:regular} 
	Counterexamples for both discrete and continuous PDFs have been devised, 
    for which regular MRSQs are strictly suboptimal~\cite{Effros:ConvexCellSuboptimal,Antos:ConvexCellSuboptimal}.
    However, none such are known for the case of log-concave input PDFs~\cite{Antos:PrivateCom}. 
    Furthermore, we shall see that such quantizers become optimal in the limit of high rates in \secref{ss:Bennett}.
\end{remark}

Our goal here is to design an MRSQ that minimizes the weighted time-average squared quantization error $\oD$ of an input $W$ with a given PDF $\fw(W)$ 
and positive weights $\{ \tG_t \}$:
\col{}{\vspace{-1.\baselineskip}}
\begin{subequations}
\label{eq:SR_cost}
\noeqref{eq:SR_cost:sum,eq:SR_cost:Dt}
\begin{align}
	\oD &= \sum_{t=1}^T \tG_t D_t ,
\label{eq:SR_cost:sum}
 \\ D_t &\triangleq \E{\left(W - \hW_t \right)^2} .
\label{eq:SR_cost:Dt}
\end{align}
\end{subequations}

Unfortunately, greedy-optimal quantizers are not globally optimal in general~\cite{DumitrescuWu:ScalarSuccessiveRefinement,Fu:GreedySuboptimal}, 
since there might be a tension between optimizing $D_{t_1}$ and $D_{t_2}$ for $t_1 \neq t_2$.
When such a tension does not exist, the source is said to be \textit{successively refinable}~\cite{EquitzCover91}, \cite[Ch.~13.5.13]{ElGamalKimBook}.

We next present a \emph{Generalized Lloyd--Max Algorithm} due to Brunk and Farvardin~\cite{GeneralizedLloydMax:SR} for constructing MRSQs, which is in turn an adaptation of an algorithm for scalar multiple descriptions by Vaishampayan~\cite{GeneralizedLloydMax:MD}.
Similarly to the standard Lloyd--Max algorithm (\algoref{algo:LM}), the generalized variant iterates between structuring the reproduction point sets $c^T$ given the partition $p_T$ (recall \remref{rem:MRSQ:partitions}), and vice versa.

Furthermore, the centroid condition of \propref{prop:Centroid} remains unaltered, as it does not have any direct effect on other stages, and is calculated separately for each stage. The partition of earlier stages, on the other hand, has a direct effect on the boundaries of newer stages, due to the nesting property~\eqref{eq:def:MRSQ:p}. 
Consequently, the nearest neighbor condition of \propref{prop:NN} is replaced by a weighted variant \cite{GeneralizedLloydMax:SR,GeneralizedLloydMax:MD}.

\begin{prop}[Weighted nearest neighbor]
\label{prop:NN:weighted}
	The optimal partition $p_T$ for a given sequence of reproduction-point sets $c^T$ is determined by the weighted nearest neighbor condition:
    \begin{subequations}
    \label{eq:intervals}
    \noeqref{eq:interval_lower,eq:interval_upper}
 	\begin{align}
    p_T [\ell] &= \max_{0 \leq i \leq 2^{\sum_{t=1}^T \Rate_t} - 1:\alpha_i < \alpha_\ell}  \frac{\beta_\ell - \beta_i}{2(\alpha_\ell - \alpha_i)} \label{eq:interval_lower}\\
    p_T [\ell+1] &= \min_{0 \leq i \leq 2^{\sum_{t=1}^T \Rate_t} - 1:\alpha_i > \alpha_\ell}  \frac{\beta_\ell - \beta_i}{2(\alpha_\ell - \alpha_i)} \label{eq:interval_upper} 
    \end{align}
    for $0 \leq \ell \leq 2^{\sum_{t=1}^T \Rate_t}-1$, where 
    \begin{align}
        m_t[\ell] & \triangleq \left\lceil (\ell+1) 2^{-\sum_{j=t+1}^T \Rate_j} \right\rceil - 1, \\
        \alpha_\ell & \triangleq \sum_{t=1}^T \tG_t c_t[  m_t[\ell] ] , 
        \\ \beta_\ell &\triangleq \sum_{t=1}^T \tG_t c^2_t[ m_t[\ell] ] .
    \end{align}
    \end{subequations}
\end{prop}

\begin{remark}
	$\alpha_\ell$ and $\beta_\ell$ can be viewed as weighted centroid and squared centroid, respectively. 
    In these terms, the partition points in \eqref{eq:interval_lower} and \eqref{eq:interval_upper} reduce to 
  the midpoints of adjacent centroids of the standard Lloyd--Max algorithm~\eqref{eq:3_opt_p}.
\end{remark}

Similarly to the optimal one-stage quantizer of \secref{ss:LM:quantizer}, 
the optimal MRSQ has to satisfy both the centroid condition of \propref{prop:Centroid} 
and the weighted nearest neighbor condition of \propref{prop:NN:weighted}, simultaneously.
Furthermore, iterating between these conditions gives rise to the Generalized Lloyd--Max algorithm.

\begin{algo}[Generalized Lloyd--Max]
\label{algo:GLM}
\

  \textbf{\textit{Initial step}.}
  Pick an initial partition $p_T$.
  
  \textbf{\textit{Iterative step}.} Repeat the two steps
    \begin{enumerate}
      \item Fix $p_T$ and evaluate $c^T$ as in \eqref{eq:3_opt_c},
      \item Fix $c^T$ and evaluate $p_T$ as in \eqref{eq:intervals},
    \end{enumerate}
    interchangeably, until the decrease in the weighted distortion $\oD$ is below a desired accuracy threshold.
\end{algo}

As in the standard Lloyd--Max algorithm, \algoref{algo:GLM} may converge to different local minima for  different initializations $p_T$. And similarly, sufficient conditions can be derived for the existence of a unique local---and thus also global---minimum~\cite{DumitrescuWu:ScalarSuccessiveRefinement}. Log-concave PDFs satisfy these conditions, suggesting that~\algoref{algo:GLM} is globally optimal for such PDFs.

\begin{thm}[\!\!\cite{DumitrescuWu:ScalarSuccessiveRefinement}]
\label{thm:GLM:optimality}
  Let the input PDF $\fw$ be log-concave and $\{\tG_t\}$---a positive weight sequence.
  Then, the Generalized Lloyd--Max algorithm converges to a unique solution that minimizes the weighted mean square error distortion~\eqref{eq:SR_cost} with weights~$\{\tG_t\}$.
\end{thm}


\subsection{Controller Design}
\label{ss:LQR:controller}

By \lemref{lem:LQG:separation}, in order to construct a globally optimal control policy, we need to find a quantizer that minimizes 
\begin{align}
\label{eq:LQR:distortion}
	\sum_{t=1}^T g_t \E{( X_t - \hX_t )^2 }.
\end{align}
The following simple result connects this problem with that of designing an MRSQ that minimizes \eqref{eq:SR_cost}.

\begin{lemma}
\label{col:MRSQ-LQR:equivalence}
    Let $\hW_t$ be the quantized description of the source sample $W_0$ at time $t \in \mT$, produced by the MRSQ that minimizes \eqref{eq:SR_cost} with weights
    \begin{align}
    \label{eq:weight}
    	\tG_t = a^{2(t-1)} g_t \,.
	\end{align}
    Then, the estimate $\hX_t$ that minimizes \eqref{eq:LQR:distortion} is given by 
    \begin{align}
    \label{eq:w2x}
     	\hX_t &= a \hX_{t-1} + U_{t-1} + a^{t-1} \left( \hW_t - \hW_{t-1} \right) , 
    \end{align}
    with $U_t = -k_t \hX_t$ and $k_t$ given in~\eqref{eq:lem:LQG:Lt}.
\end{lemma}

\begin{IEEEproof} 
	Recall that $X_t$ is given by the recursion
    \begin{subequations}
    \label{eq:system:LQR}
    \noeqref{eq:system:LQR:t=0}
    \begin{align}
    	X_{t+1} &= a X_t + U_t , & t \in \interval{T - 1} ,
    \label{eq:system:LQR:general}
     \\ X_1 &= W_0 \,.
    \label{eq:system:LQR:t=0}
    \end{align}
    \end{subequations}
    The corresponding explicit expression for $X_t$ in this case is
    \begin{align}
    \label{eq:LQR:x:explicit}
    	X_t =  a^{t-1}  W_0 + \sum_{i = 0}^{t-1}  a^{t - 1 - i} U_i \,.
    \end{align}
    This suggests, in turn, that the estimate $\hX_t$ of $X_t$ at time $t$ can be expressed as 
    \begin{subequations}
    \label{eq:w2x:proof}
    \noeqref{eq:w2x:proof:def_hx,eq:w2x:proof:sub,eq:w2x:proof:def_hw,eq:w2x:proof:explicit,eq:w2x:proof:recur}
    \begin{align}
       \hX_t &= \CE{X_t}{\PACKET^t}
    \label{eq:w2x:proof:def_hx}
     \\ &= \CE{a^{t-1} W_0 + \sum_{i = 0}^{t-1} a^{t - 1 - i} U_i}{\PACKET^t}
    \label{eq:w2x:proof:sub}
     \\ &= a^{t-1}  \left( \hW_t + \sum_{i = 0}^{t-1} a^{-i} U_i \right)
    \label{eq:w2x:proof:def_hw}
     \\ &= a^{t-1} \left( \hW_t - \hW_{t-1} \right) + U_{t-1} 
     \\ &\qquad\qquad + a \cdot a^{t-2} \left( \hW_{t-1} + \sum_{i = 0}^{t-2} a^{-i} U_i \right)
    \label{eq:w2x:proof:explicit}
     \\ &= a^{t-1} \left( \hW_t - \hW_{t-1} \right) + U_{t-1} + a \hX_{t-1} \,,
    \label{eq:w2x:proof:recur}
    \end{align}
    \end{subequations}
    which proves the relation in \eqref{eq:w2x}, 
    where \eqref{eq:w2x:proof:def_hx} follows from the definition of $\hX_t$, 
    \eqref{eq:w2x:proof:sub} holds due to \eqref{eq:LQR:x:explicit},
    \eqref{eq:w2x:proof:def_hw} follows from the definition of $\hW_t$ and the action $U_t$ being a function of $\PACKET^t$, 
    and \eqref{eq:w2x:proof:recur} holds by substituting the relation established in \eqref{eq:w2x:proof:def_hw} for $\hX_{t-1}$, namely, 
    \begin{align}
    \label{eq:LQR:hx:explicit}
    	\hX_{t-1} = a^{t-2}  \left( \hW_t + \sum_{i = 0}^{t-2}  a^{-i} U_i \right) .
    \end{align}

	Subtracting \eqref{eq:LQR:hx:explicit} from \eqref{eq:system:LQR:general} with the appropriate index adjustments concludes the proof.
\end{IEEEproof}

We are now ready to present the globally optimal control policy for the LQR problem.

\begin{algo}[Globally optimal LQR control]
\label{algo:LQR}
\

    \textbf{\textit{Initialization}.} 
    Both the encoder and the decoder:
    \begin{enumerate}
    \item
    	Construct $\{s_t, k_t, g_t | t \in \mT \}$ as in \lemref{lem:LQG:separation} 
    	for the given $T$, $\{\CostXs_t\}$, $\{\CostUs_t\}$ and $a$.
	\item
    	Set $\tG_t = A^{2(t-1)} g_t$ as in \eqref{eq:weight}.
	\item
    	Construct the $T$-step MRSQ sequence $\Quantizer^T$ using \algoref{algo:GLM} for the source $W_0$ and weights $\{\tG_t\}$.
    \end{enumerate}

    \textbf{\textit{Observer/Encoder}.}
	Observes $w_0$.
    At time $t \in \interval{T-1}$:
    \begin{enumerate}
    \item
    	Generates the quantizer index: $\packet_t = \mE_{\Quantizer_t} \left( w_0 \right)$.
	\item
    	Transmits $\packet_t$.
    \end{enumerate}
    \textbf{\textit{Controller/Decoder}.} At each time $t \in \interval{T-1}$:
    \begin{enumerate}
    \item
    	Receives $\packet_t$.
	\item
    	Generates the description: $\hw_t = \mD_{\Quantizer_t}( \packet^t ) 
        = \Quantizer_t( w_0 )$.
    \item
    	Generates $\hx_t$ as in \eqref{eq:w2x}.
    \item
    	Generates the control actuation $u_t = - k_t \hx_t$.
    \end{enumerate}
\end{algo}
Combining Lemmata \ref{lem:LQG:separation} and \ref{col:MRSQ-LQR:equivalence} with \thmref{thm:GLM:optimality} leads to the global optimality of Algorithm \ref{algo:LQR}.

\begin{thm}
\label{thm:LQR_opt}
	Let $\fw$ be a log-concave PDF. Then, \algoref{algo:LQR} achieves the minimum possible average-stage LQ cost~\eqref{cost}.
\end{thm}


\subsection{High-Rate Limit}
\label{ss:Bennett}

We now consider the high resolution case, 
\viz\ the case in which the rates $R^T$ are large.\footnote{The exact notion of a large rate will become clear in the sequel.}
We start by treating the case of a single rate ($T=1$).

We follow the exposition in~\cite[Ch.~6.3]{GershoGrayBook} of Bennett's approximated quantization law for a single target rate $R$ (recall \secref{ss:LM:quantizer}).

For a large enough rate $R$, 
and consequently small enough cell widths $\{p[\packet+1] - p[\packet]\}$ (except for maybe the extreme cells), 
the sum in \eqref{eq:dist-measure:regular} can be approximated by a Riemann integral 
by defining a reproduction-point PDF
\begin{align}
	\nu(x) \triangleq \lim_{N \to \infty} \frac{N(x)}{N} ,
\end{align}
where $N = 2^R$ is the number of reproduction points $c = \{ c_0, \ldots, c_{N-1} \}$, 
and $N(x) \Delta x$ is the approximate number of points in $[x, x + \Delta x)$ for a small $\Delta x$.
In this limit, the size of cell $i$ is approximated by 
\begin{align}
\label{eq:Bennett:interval}
	p[\ell+1] - p[\ell] \approx \frac{1}{N \nu(c_\ell)} \,.
\end{align}

\begin{thm}[Bennett's law]
	The optimal reproduction-point PDF $\nu$ of a source with a log-concave PDF $\fw$,\footnote{This holds true for a much wider class of sources with smooth PDFs~\cite{GrayNeuhoff:Quantization:IT1998}.}
    in the limit of large $N$, 
    is given by 
    \begin{align}
\label{eq:Bennett:PDF}
    	\nu(x) \approx \frac{\fw^{1/3}(x)}{ \int\limits_{\xi \in \supp{f}} \fw^{1/3}(\xi) d \xi} \,,
    \end{align}
    and achieves a distortion~\eqref{eq:dist-measure} 
    \begin{align}
    \label{eq:Bennett:D}
        D &\approx \frac{1}{12 \times 2^{2R}} \left[ \int\limits_{x \in \supp{\fw}} \fw^{1/3}(x) dx \right]^3 .
    \end{align}
\end{thm}

An immediate consequence of this theorem is that, 
in the limit of high rates ($\Rate_1 \gg 1$),
the source is successively refinable, 
since the approximation of \eqref{eq:Bennett:interval} and \eqref{eq:Bennett:PDF} 
is tightened by each subsequent additional rate~\cite[Thm.~5]{DumitrescuWu:ScalarSuccessiveRefinement}.

\begin{corol}[Succesive refinability]
	A source with a log-concave PDF $\fw$ is approximately successively refinable 
    in the limit of large $R_1$ (and consequently large $\sum_{i = 1}^t \Rate_i$), 
    with a reproduction-point PDF as in~\eqref{eq:Bennett:PDF}, 
    and distortions $D^T$ with $D_t$ given as in~\eqref{eq:Bennett:D} with rate $\sum_{i = 1}^t \Rate_i$.
\end{corol}

\begin{remark}
	Bennett's law holds true for a wider class of PDFs and distortions;
    see \secref{ss:discuss:cost} for further discussion.
\end{remark}


\section{Numerical Calculations}
\label{s:numeric}

\subsection{Greedy LQG Control}
\label{numeric:LQG}

\begin{figure}[tp]
\vspace{.5\baselineskip}
\centering
	\col{\includegraphics[width=.65\columnwidth]{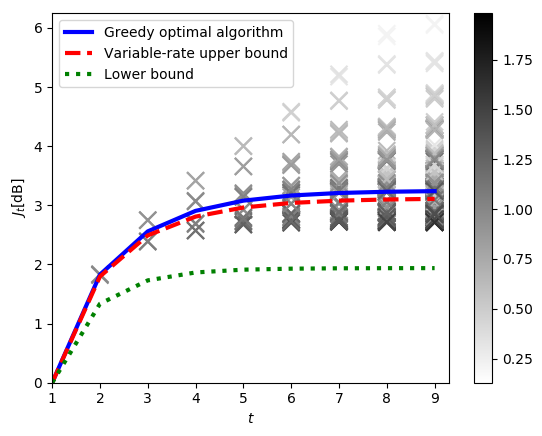}}{\includegraphics[width=\columnwidth]{./fig2_drop_dB}}
\vspace{-\baselineskip}
    \caption{Instantaneous LQ control cost $\LQGcost_t$ as a function of the time $t$ over a channel of an optimal greedy algorithm with $a = 1.2$, $\VarW=1$, $\Rate_t \equiv 1$ along with upper and lower bounds for the rate-variable case. The 'x' marks correspond to instantaneous costs $\LQGcost_t$ for $2^{t\Rate}$ contiguous intervals at time $t$ and the intensity of gray scale for $\LQGcost_t$ for each interval indicates the relative magnitude of probability of falling into that interval at time $t$.}
    \label{fig:subcell}
\end{figure}

We now evaluate the instantaneous costs~\eqref{eq:cost:instant} of \algoref{algo:greedy} 
for a standard Gaussian i.i.d.\ disturbance sequence $\{W_t\}$, $\CostXs_t \equiv 1$, $\CostUs_t \equiv 0$, and $a = 1.2$. These costs are depicted in~\figref{fig:subcell}
along with $\E{( \hX_t - X_t)^2| \PACKET^t = \packet^t }$ for all admissible transmit sequences $\packet^t$.
We compare them to the following upper and lower bounds, also depicted in~\figref{fig:subcell}, which are valid for the less restrictive case of variable-rate feedback~\cite[Ch.~9.9]{GershoGrayBook}, 
where the \textit{average rate across time} is constrained by $\Rate$. 

\begin{prop}[\!\!{\cite{SilvaDerpichOstergaard:ECDQ4Control,KostinaHassibi:Allerton2016,StreamingWithFB:ITW2017,StreamingWithFB:CNS}}]
\label{prop:LB}
    Consider the setting of a variable-rate subject to an expected-rate constraint $\Rate$, 
    i.i.d.\ Gaussian disturbances $\{W_t\}$ of variance $\VarW$, $\CostXs_t \equiv 1$ and $\CostUs_t \equiv 0$.
    Then, the instantaneous cost $\LQGcost_t$ is bounded as $\LQGcost^\text{LB}_t \leq \LQGcost_t \leq \LQGcost^\text{UB}_t$, with $\LQGcost^\text{UB}_0 = \LQGcost^\text{LB}_0 = 0$ and 
    \begin{align}
	    \LQGcost^\text{LB}_{t+1} &= a^2 \LQGcost^\text{LB}_t 2^{-2\Rate} + \VarW, 
	 \\
     \LQGcost^\text{UB}_{t+1} &= \frac{2\pi\e}{12} a^2 \LQGcost^\text{UB}_t 2^{-2\Rate} + \VarW.
	\nonumber
	\end{align}
\end{prop}


\subsection{LQR Control}
\label{ss:numeric:LQR}

Here we compare the performance of the optimal greedy and globally optimal algorithms 
for LQR control with $a=1.5$, $\CostXs_t \equiv 1$, $\CostUs_t \equiv 0$, standard Gaussian disturbance $W_0$, $W_t = 0$ for $t \geq 1$, 
and time horizon $T = 9$.
The accumulated costs~\eqref{cost} for $t = 1, \cdots, 9$ 
are tabulated in \tableref{table_example}. 

\begin{table}
\vspace{.7\baselineskip}
\begin{center}
    \begin{tabular}{| c | c | c |}
    \hline
    t   & Greedy & Optimal 							\\ \hline\hline
    1	&\mynum{1}& \mynum{1}						\\ \hline
    2	&\mynum{1.81757991}	&\mynum{1.81770572}		\\ \hline
    3	&\mynum{2.41254276}	&\mynum{2.41259522}		\\ \hline
    4	&\mynum{2.80988449}	&\mynum{2.80638865}		\\ \hline
    5	&\mynum{3.06143659}	&\mynum{3.05137884}		\\ \hline
    6	&\mynum{3.21559569}	&\mynum{3.20009849}		\\ \hline
    7	&\mynum{3.30792786}	&\mynum{3.28771711}		\\ \hline
    8	&\mynum{3.36235429}	&\mynum{3.33811839}		\\ \hline
    9	&\mynum{3.39407619}	&\mynum{3.36884029}		\\ \hline
    \end{tabular}
\vspace{\baselineskip}
	\caption{Optimality gap of the greedy algorithm for time horizon $T = 9$ and $A = 1.5$.}
\label{table_example}
\end{center}
\vspace{-2\baselineskip}
\end{table}

The table demonstrates that \algoref{algo:LQR} is globally optimal for a predefined specific time horizon ($T=9$ in the example) and might be suboptimal for other time instants. Since \algoref{algo:greedy} optimizes the performance in a greedy fashion it always achieves the optimal instantaneous cost at time $t=2$, $\LQGcost_2$;\footnote{At time $t=1$, the cost $\LQGcost_1$ is the same regardless of the algorithm applied.}
for a time horizon $T > 2$ this would mean superior performance over the \algoref{algo:LQR} as is evident by the second row in \tableref{table_example}.


\subsection{Event-Triggered Control}
\label{ss:numeric:event-trig}

\begin{figure}[tp]
\centering
	\col{\includegraphics[width=.65\columnwidth, trim = {5mm 7mm 4mm 7mm}]{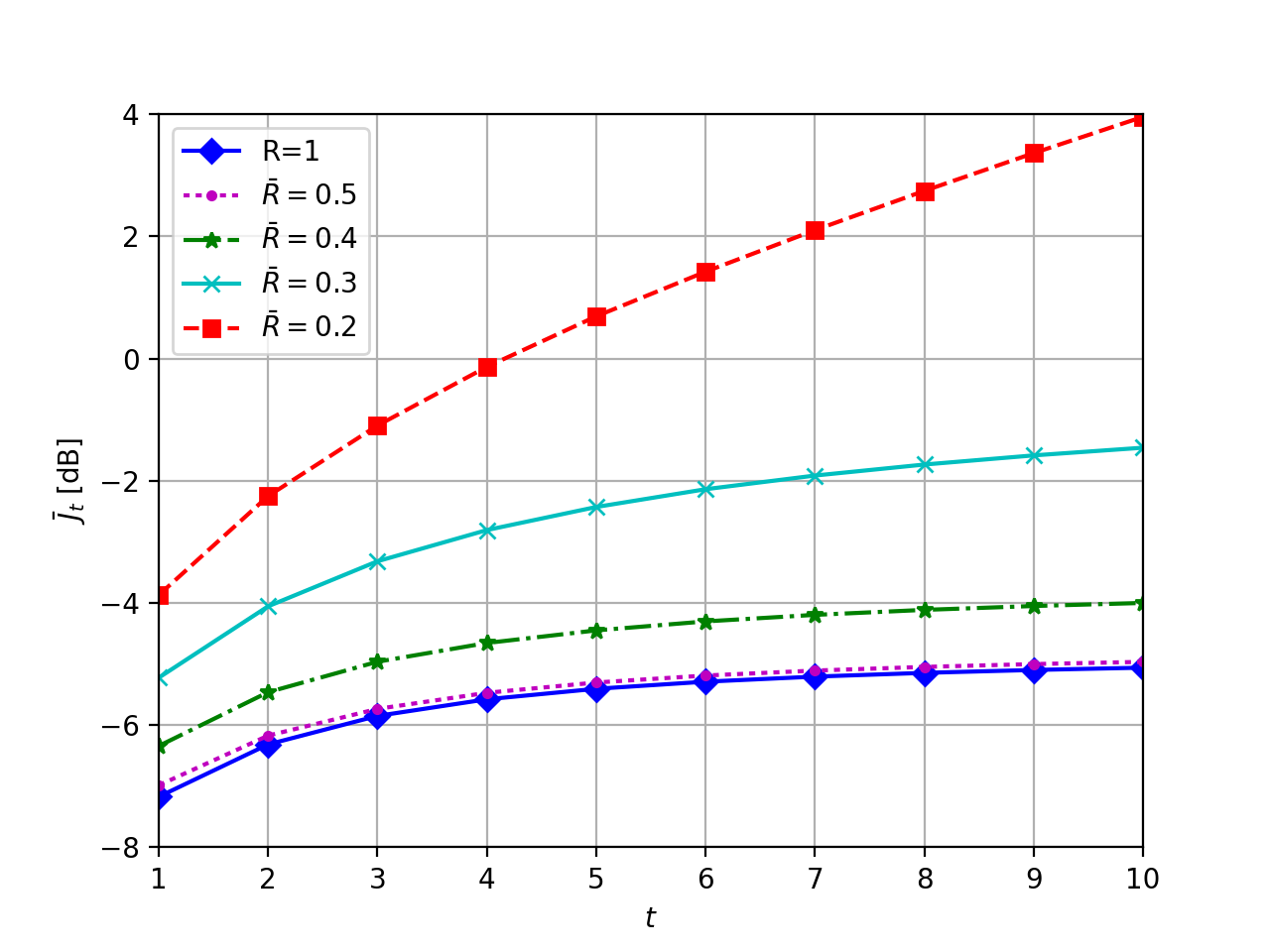}}{\includegraphics[width=\columnwidth, trim = {5mm 3mm 4mm 7mm}]{./event_trig_perf}}
    \caption{time-average LQ cost $\oLQGcost$ as a function of the time $t$ of event-triggered control with different average-rate $\oR$ values, and time-triggered control of $\Rate = 1$, with $a = 1.5, \VarW = 1, \CostXs_t \equiv 1, \CostUs_t \equiv 0$.}
    \label{fig:event-triggered}
\end{figure}

We compare in \figref{fig:event-triggered} the performance of \algoref{algo:greedy} of rate $R = 1$ with the event-triggered algorithm (\algoref{algo:LM_silence}) for various transmission rates $\oR$, 
for the LQG setup with $a=1.5$, $\CostXs_t \equiv 1$, $\CostUs_t \equiv 0$, and i.i.d.\ standard Gaussian disturbance ($\VarW=1$).

Note that \algoref{algo:greedy} of rate $R=1$ is equivalent to \algoref{algo:LM_silence} without any constraint on $\oR$, \ie, $\oR = 1$.

\section{Discussion and Future Work}
\label{s:discuss}


\subsection{Global Optimality}
\label{ss:discuss:global-opt}

As is seen in \secref{ss:numeric:LQR}, 
even in the extreme case of low rate $\Rate_t \equiv 1$, 
the improvement of the globally optimal algorithm is negligible over the the optimal greedy algorithm is negligible (fractions of a percent)---a fact previously observed in~\cite{Fu:GreedySuboptimal}. 
For high rates, the gap is negligible as is evident from~\secref{ss:Bennett}.
This seems to extend to the more general case of i.i.d.\ disturbances (LQG case included), 
as adding an independent noise can only reduce the gap between the two quantizers, 
suggesting that the optimal greedy algorithm is essentially optimal for all practical purposes.

An interesting avenue would be to explore an even lower-complexity algorithm. 
A noteworthy attempt was made by Y\"uksel who considered a low-complexity uniform adaptive quantizer. 
Unfortunately, such quantizers, being inherently symmetric, 
cannot stabilize any unstable system using one-bit quantization rate, 
as no zooming in/out is possible this way. 
This is in stark contrast to the Lloyd--Max based algorithms that can become non-symmetric even for a rate of one bit, 
via repeated convolution of same size PDF tails; this is evident from \secref{s:numeric}.


\subsection{Other Cost Metrics}
\label{ss:discuss:cost}

We concentrated on optimizing the quadratic cost~\eqref{cost} in this work.
However, the results of this work extend to any strictly-increasing,
convex and continuously-differentiable difference cost measure~\cite{TrushkinConditions:Kieffer,TrushkinConditions:Trushkin,DumitrescuWu:ScalarSuccessiveRefinement} in a straightforward fashion, 
since the guarantees for the Lloyd--Max and Generalized Lloyd--Max algorithms hold true for such cost measures (and log-concave PDFs); note that this includes all $k$\textsuperscript{th}-moment distortion measures with $k \geq 1$. 

Furthermore, in the high-resolution regime of \secref{ss:Bennett}, Bennett's law holds for a much wider class of PDFs and can be easily adapted for the more general case of $k$\textsuperscript{th}-moment distortion measures with $k \geq 1$~\cite{GrayNeuhoff:Quantization:IT1998}, \cite[Th.~5]{DumitrescuWu:ScalarSuccessiveRefinement}.


\subsection{Optimal Minimal Cell-probability Constrained Quantization}
\label{ss:discuss:conv}

(Unconstrained) Lloyd--Max quantization (\algoref{algo:LM}) 
is guaranteed to converge to the global optimum for log-concave PDFs. 
It would be interesting to prove a similar result for 
its constrained variant---\algoref{algo:LM_silence}.
We verified this numerically for Gaussian, exponential and Laplace PDFs and conjecture that it holds true for all log-concave PDFs.


\subsection{Partially-observed Systems}
\label{ss:discuss:cost}

In this paper, we concentrated on the simplest NCS scenario---a scalar linear fully-observable system.
However, \assertref{assert:LogConcaveProps}
suggests that the results of this work can be extended\col{}{

\noindent} to the case of partially-observed linear systems as long as the measured outputs are scalar; this is left for future work.

For non-scalar measurements (and systems), 
scalar quantization needs to be replaced with vector quantization~\cite[Ch.~3]{GershoGrayBook}. 
Unfortunately, in this case, Lloyd--Max quantization does not converge to a unique solution for different initializations for log-concave (and even i.i.d.\ Gaussian) PDFs~\cite[Ch.~III.11]{GershoGrayBook}.
Nonetheless, the gain of vector quantization over scalar (per-dimension) quantization is bounded~\cite[Ch.~9]{ZamirBook}. 
Finally, we note that Bennett's high-resolution law of \secref{ss:Bennett} extends straightforwardly to the multi-dimensional case.


\subsection{Delayed Arrivals and Acknowledgments}
\label{ss:discuss:global-opt}

The results in this work include the case of packet erasures when knowledge of the erasures is made available to the observer/encoder prior to the transmission of the next packet. 

The extension to the case of delayed acknowledgments was considered in~\cite{StreamingWithFB:ITW2017,StreamingWithFB:CNS} for the case of variable-rate feedback; studying this problem in the fixed-rate feedback scenario remains an interesting avenue to explore. 
Similar ideas seem to extend also to the case of delayed packet arrivals, 
and are left for future research.



\section*{Acknowledgment}

A.~Khina thanks M.~J.~Khojasteh and M.~Franceschetti for many stimulating and helpful discussions, and especially for introducing him to event-triggered control and pointing his attention to
\cite{KofmanBraslavsky:CDC2006,khojasteh:Allerton2016,khojasteh:CDC2017,EventTriggered:PearsonHespanhaLiberzon}.


\bibliographystyle{IEEEtran}


\end{document}